\documentclass[12pt,thmsa]{article}
\usepackage{sw20lart}

\input{tcilatex}
\begin{document}

\author{Liping Fu, Jun Luo, Li Xiao and Xizhi Zeng \\
\textit{\small Laboratory of Magnetic Resonance and Atomic and Molecular
Physics,\ }\\
\textit{\small \ Wuhan Institute of Physics and Mathematics, The Chinese\
Academy of Sciences, }\\
\textit{\small \ \ Wuhan 430071, People's Republic of China }}
\title{Experimental Realization of Discrete Fourier Transformation on NMR Quantum
Computer }
\date{5/15/99/ }
\maketitle

\begin{abstract}
We report experimental implementation of discrete Fourier
transformation(DFT) on a nuclear magnetic resonance(NMR) quantum computer.
Experimental results agree with theoretical results. Using the pulse
sequences we introduced, DFT can be realized on any $L$-bit quantum number
in principle.\medskip \newline
\rule{0in}{0in}\medskip PACS: 87.70.+c, 03.65.-w\newline
\emph{keywords: }NMR; discrete Fourier tranform; experiment; simulation;
Shor's algorithm
\end{abstract}

\section{INTRODUCTION}

With the discovery of an apparent separation between the classical and
quantum classifications of computational complexity\cite{shor1},and of
fault-tolerant schemes for quantum computation\cite{shor2}, quantum
information theory plays an important role in computer science. In 1994, Shor%
\cite{shor1}introduced the quantum factoring algorithm, which achieves an
exponential speed-up relative to classical algorithms. Shor's crucial
insight was that the discrete Fourier transformation(DFT) can be evaluated
in polynomial time on a quantum computer. In order to realize Shor's
algorithm, performing DFT becomes most important. In this paper, we
introduce a version to realize DFT by using NMR spectrometer and simulator
and give our experiment and simulation results realizing two qubit DFT.

The discrete Fourier transformation modulo$q$(DFT$_{q}$) is a unitary
transformation in $q$-dimension($q=2^{L}$). It is defined relative to chosen
basis $\mid 0\rangle ,\cdot \cdot \cdot ,\mid q-1\rangle $ by\cite{Ekert}:

DFT$_{q}:\mid a\rangle \longrightarrow \frac{1}{\sqrt{q}}\stackrel{q-1}{%
\stackunder{c=0}{\sum }}\exp (2\pi iac/q)\mid c\rangle $ \qquad \quad \qquad
\qquad \qquad $(1)$

Coppersmith\cite{Copp} suggested the efficient algorithm for DFT on
application of quantum mechanical operators and gave an improvement in which
evaluation of $L$ bit Fourier transformation is accomplished by composing $L$
one-qubit operations and $\frac{1}{2}L(L-1)$ two-qubit operations. So the
implementation of DFT is based on realizing one-qubit operator $A_{j}$ which
acts on the state of the $j$-qubit and two-qubit operator $B_{jk}$ which
acts on the states of qubit $j$ and $k$. The explicit form of the operator $%
A_{j}$ is\cite{Ekert}:

$A_{j}=2^{-1/2}(\mid 0_{j}\rangle \langle 0_{j}\mid +\mid 0_{j}\rangle
\langle 1_{j}\mid +\mid 1_{j}\rangle \langle 0_{j}\mid -\mid 1_{j}\rangle
\langle 1_{j}\mid )$ \qquad \qquad $(2)$

\qquad $\qquad \qquad (j=0\cdot \cdot \cdot \cdot \cdot \cdot )$

The matrix representation of $A_{j}$ is: $\frac{1}{\sqrt{2}}\left( 
\begin{array}{ll}
1 & 1 \\ 
1 & -1
\end{array}
\right) $

The explicit form of the operator $B_{jk}$ is\cite{Ekert}:

$B_{jk}=\mid 0_{j}0_{k}\rangle \langle 0_{j}0_{k}\mid +\mid
0_{j}1_{k}\rangle \langle 0_{j}1_{k}\mid +\mid 1_{j}0_{k}\rangle \langle
1_{j}0_{k}\mid +e^{i\theta _{jk}}\mid 1_{j}1_{k}\rangle \langle
1_{j}1_{k}\mid $ $(3)$\qquad \qquad \ \qquad \qquad

\qquad \qquad \qquad \qquad \qquad $(\theta _{jk}=\frac{\pi }{2^{k-j}})$

The matrix representation of $B_{jk}$ is: $\left( 
\begin{array}{llll}
1 & 0 & 0 & 0 \\ 
0 & 1 & 0 & 0 \\ 
0 & 0 & 1 & 0 \\ 
0 & 0 & 0 & e^{i\theta }
\end{array}
\right) .$

Transformation of $B_{jk}$ affects only state $\mid 1_{j}1_{k}\rangle .$

We use a network to illustrate how to use operators $A_{j}$ and $B_{jk}$ to
compose $5$-qubit transformation in $\emph{Figure.}$ $\emph{1}$. The
operators corresponding to the network are:

$%
(A_{a})(B_{34}A_{3})(B_{24}B_{23}A_{2})(B_{14}B_{13}B_{12}A_{1})(B_{04}B_{03}B_{02}B_{01}A_{0}) 
$

\section{IMPLEMENTING THE DFT IN NMR}

In this paper we realize DFT by using NMR. The prime thing we should do is
to perform $A_{j}$ and $B_{jk}$ by using radiofrequency pulses and spin-spin
interaction. Using NMR to perform quantum computer, we should choose AX
couple system. The Hamiltonian for this system can be approximated as:

$H=2\varpi _{AB}J_{ZA}J_{ZB}+\varpi _{A}J_{ZA}+\varpi _{B}J_{ZB}+H_{env}$
\qquad \qquad \qquad \qquad \qquad (4)

We denote the rotation $\varphi _{X}^{j(k)}=\exp (i\varphi I_{x})$ for a $%
\varphi $ rotation about the $\stackrel{\wedge }{x}$-axis and $\varphi
_{Y}^{j(k)}=\exp (i\varphi I_{y})$ for a $\varphi $ rotation about the $%
\stackrel{\wedge }{y}$-axis, $j$ or $k$ means which nucleus is operated on,
and $\tau $ is the time within which the system undergoes the unitary
transformation $\exp (2\pi iJ_{ZA}J_{ZB}t)$ in the doubly rotating frame.

To perform $A_{j}$ is actually to perform a Walsh-Hadamard transformation,
which rotates each quantum qubit from $\mid 0\rangle $ to $(\mid 0\rangle
+\mid 1\rangle )/\sqrt{2}$. the pulse sequences is\cite{Chuang}: $(\frac{\pi 
}{2})_{X}-(\frac{\pi }{2})_{X}-(-\frac{\pi }{2})_{Y}$

To perform $B_{jk},$ we use radiofrequency pulses and spin-spin interaction
like:

$(-\frac{\pi }{2})_{Y}^{j}(-\frac{\pi }{2})_{Y}^{k}-(-\varphi
)_{X}^{j}(-\varphi )_{X}^{k}-(\frac{\pi }{2})_{Y}^{j}(\frac{\pi }{2}%
)_{Y}^{k}-\frac{\tau }{2}-(\pi )_{X}^{j}(\pi )_{X}^{k}-\frac{\tau }{2}-(\pi
)_{X}^{j}(\pi )_{X}^{k}$

\qquad \qquad \qquad $(\varphi =\frac{\pi }{2^{k-j+1}},\tau =\frac{1}{%
J*2^{k-j+1}})$

Using the pulse sequences and spin-spin interaction we introduced above, we
can implement DFT to any quantum number with $L$-qubit in principle.

\section{EXPERIMENTAL AND SIMULATION RESULTS}

In order to demonstrate the results described above, we have constructed an
NMR quantum computer capable of implementing the DFT. In this section, we
give NMR experimental and NMR simulation results after doing DFT on state $%
\mid 0_{j}1_{k}\rangle $ by using the two radiofrequency pulse sequences
corresponding to the operator $A_{j}$ and $B_{jk}$ respectively. Data of the
NMR experiment were taken at room temperature with a Bruker ARX-500
spectrometer and the signals were obtained with single-shot measurement.%
\emph{\ }

In the experiment we chose $H_{2}PO_{3}$ as our sample, labeled $^{31}P$ as $%
j$-qubit and $^{1}H$ as $k$-qubit. The observed J-coupling between $^{1}H$
and $^{31}P$ was $647.451Hz.$ First, we produced effective pure state $\mid
00\rangle $by using ''temporal averaging'\cite{Chuang1}. The pulses we used
were\cite{John}:

\emph{E: NONE}

\emph{P1: }$(\frac{\pi }{2})_{X}^{p}-\frac{1}{2}J-(\frac{\pi }{2})_{Y}^{p}(%
\frac{\pi }{2})_{X}^{H}-\frac{1}{2J}-(\frac{\pi }{2})_{Y}^{H}$

\emph{P2: }$(\frac{\pi }{2})_{X}^{H}-\frac{1}{2}J-(\frac{\pi }{2})_{Y}^{H}(%
\frac{\pi }{2})_{X}^{P}-\frac{1}{2J}-(\frac{\pi }{2})_{Y}^{P}$

Since we wanted to perform DFT on the state$\mid 01\rangle ,$ so after the
preparation of the state $\mid 00\rangle ,$we should operate the pulse
sequence: $(\frac{\pi }{2})_{Y}^{H}-\frac{1}{2J}-(\frac{\pi }{2})_{X}^{H}$
on $^{1}H$ to obtain the state$\mid 01\rangle .$

Second, we performed discrete Fourier transformation on $\mid 01\rangle .$
We list the pulses and spin-spin interaction are as follows:

$(\pi )_{X}^{P}-(-\frac{\pi }{2})_{Y}^{p}-(\frac{\pi }{2})_{Y}^{p}(\frac{\pi 
}{2})_{Y}^{H}-(\frac{\pi }{4})_{X}^{p}(\frac{\pi }{4})_{X}^{H}-(-\frac{\pi }{%
2})_{Y}^{p}(-\frac{\pi }{2})_{X}^{H}-\frac{1}{4J}-(\pi )_{X}^{p}-(-\frac{\pi 
}{2})_{X}^{H}$

The first two pulses and the last two pulses correspond to the operator $%
A_{j}$ and $A_{k}$ respectively, the rest pulses correspond to $B_{jk}.$
After these operations, the initial state$\mid 01\rangle $ was transformed
into state:

$\frac{1}{2}$ $(\mid 00\rangle -\mid 01\rangle +i\mid 10\rangle -i\mid
11\rangle .$ \qquad \qquad \qquad \qquad \qquad $(5)$

Compared with the definition of DFT $(1)$ which means that after doing DFT
operation, the state$\mid 01\rangle $ should become

$\mid 01\rangle \longrightarrow \frac{1}{2}$ $(\mid 00\rangle +e^{i\frac{\pi 
}{2}}\mid 01\rangle +e^{i\pi }\mid 10\rangle +e^{i\frac{3\pi }{2}}\mid
11\rangle )$

$\qquad \qquad =\frac{1}{2}$ $(\mid 00\rangle +i\mid 01\rangle -\mid
10\rangle -i\mid 11\rangle )$ \qquad \qquad \qquad \qquad $(6)$

and the density matrix is: $\left( 
\begin{array}{llll}
1 & -i & -1 & i \\ 
i & 1 & -i & -1 \\ 
-1 & i & 1 & -i \\ 
i & -1 & i & 1
\end{array}
\right) ,$\qquad \qquad \qquad \qquad $(7)$

The third step: we should reverse the qubit. That is, for example, for the
case of three qubits, $\mid ijk\rangle \longrightarrow \mid kji\rangle .$
The third step is composed of three C-NOT operations, The network is shown
in $\emph{Figure.}$ $\emph{2.}$ The radiofrequency pulses corresponding to
the three C-NOT operators are:

1. $(\frac{\pi }{2})_{Y}^{H}-\frac{1}{2J}-(-\frac{\pi }{2})_{Y}^{P}(-\frac{%
\pi }{2})_{Y}^{H}-(-\frac{\pi }{2})_{X}^{P}(\frac{\pi }{2})_{X}^{H}-(\frac{%
\pi }{2})_{Y}^{P}$

2. $(\frac{\pi }{2})_{Y}^{P}-\frac{1}{2J}-(-\frac{\pi }{2})_{Y}^{P}(-\frac{%
\pi }{2})_{Y}^{H}-(\frac{\pi }{2})_{X}^{P}(-\frac{\pi }{2})_{X}^{H}-(\frac{%
\pi }{2})_{Y}^{H}$

3. $(\frac{\pi }{2})_{Y}^{H}-\frac{1}{2J}-(-\frac{\pi }{2})_{Y}^{P}(-\frac{%
\pi }{2})_{Y}^{H}-(-\frac{\pi }{2})_{X}^{P}(\frac{\pi }{2})_{X}^{H}-(\frac{%
\pi }{2})_{Y}^{P}$

Actually, the quantum-mechanical operation ''reverse the qubit'' is not
applied. Instead of this, one measures the state after DFT and reads the
result of the measurement in the opposite order\cite{Berman}.

After these three steps, we had finished the operation of DFT on the initial
state $\mid 01\rangle $ . In order to illustrate the result we got, we need
to obtain all the elements in the two-spin density matrix of the ultimate
state by using state tomography\cite{Chuang3}. Before we did the experiment,
we had simulated the experiment by using NMR simulator. The results are
shown in $Figure$ $3a$ $,$ it can be regarded as a theoretical result. The
matrix representation of the result is: $\left( 
\begin{array}{llll}
1 & -i & -1 & i \\ 
i & 1 & -i & -1 \\ 
-1 & i & 1 & -i \\ 
-i & -1 & i & 1
\end{array}
\right) ,$ which is the same as we expected.

An advantage of doing NMR simulator is that it can represent the physics
problem in spite of considering experimental errors (chemical shift, the
affection of environmental and so on). So using NMR simulator can be
considered as a quantum computer to simulate some quantum mechanical problem.

The experimental results are shown in $Figure$ $3b$, we measured and got
real and imaginary components of the integral area from each peak, then we
used all the integral areas we'd got to reconstruct that density matrix.
Compared with the theoretical results, the experimental results agree with
those deduced from theory. The relative errors are $24.8\%$ and $4.7\%$.
They primarily due to the imperfect calibration of the rotation, and are
also caused by inhomogeneity of the magnetic field or magnetization and
least-squares fitting used in the tomography procedure. Using other NMR
techniques, such as phase cycling, the errors can be minimized.

\section{CONCLUSION}

We have demonstrated that the discrete Fourier transition can be implemented
by means of NMR spectrometer and simulator and the experiment agrees with
the theory well. Using the method we introduced, we can perform DFT on any
quantum number with $L$-qubit. Performing DFT successfully gives a
probability to realize Shor's algorithm by using NMR. Fourier analysis is a
versatile tools in the laboratory\cite{Preskill}, so we might expect that
the fast DFT should be an important application to physics.

\end{document}